\documentclass[epj-spec]{svjour}
\usepackage{graphics}
\begin{document}
\title{W-like states of $N$ uncoupled spins $\frac{1}{2}$}
\author{E. Ferraro\inst{1}, A. Napoli\inst{1}, M. A. Jivulescu \inst{1,2}, A. Messina\inst{1}}
\institute{CNISM and  Dipartimento di Scienze Fisiche ed
Astronomiche, Universit\`{a} di Palermo, via Archirafi 36, 90123
Palermo, Italy \and Department of Mathematics, "Politehnica"
University of Timi\c{s}oara, P-ta Victoriei Nr. 2, 300006
Timi\c{s}oara, Romania}

\abstract{The exact dynamics of a disordered spin star system,
describing a central spin coupled to $N$ distinguishable and non
interacting spins $\frac{1}{2}$, is reported. Exploiting their
interaction with the central single spin system, we present
possible conditional schemes for the generation of W-like states,
as well as of well-defined angular momentum states, of the $N$
uncoupled spins. We provide in addition a way  to estimate the
coupling intensity between each of the $N$ spins and the central
one. Finally the feasibility of our procedure is briefly
discussed.} \maketitle
\section{Introduction}
Interacting spin models play a central role in many physical
contexts providing a paradigm to describe a wide range of
different systems. In condensed matter physics, for example, they
can be explored to analyze many properties of magnetic compounds.
It is indeed well known that a suitable general model of a magnet
consists of $N$ spins coupled by exchange interaction with
arbitrary range and strength.

The interest toward spin systems, and more in particular toward
spin dynamics in semiconductor structures, has remarkably
increased in the last few years also in connection with the new
emerging areas of quantum computation and information
\cite{Bayat}-\cite{Plenio}. In these contexts, spin models like
Heisenberg spin chains or spin star systems, describing for
example a single electron spin in a semiconductor quantum dot
interacting with surrounding nuclear spins via hyperfine coupling
mechanisms, have been extensively studied
\cite{Pratt}-\cite{Breuer}.

Generally speaking spin models have proved to be promising
candidates for the generation and the control of assigned quantum
correlations, as witnessed by the numerous papers recently
appeared in literature\cite{Verstraete}-\cite{Binosi}.

In this paper we concentrate on the possibility of manipulating at
demand the state of a sample of $N$ uncoupled spins exploiting
their common interaction with another single spin called central
system. The analysis we have developed, on the one hand provides
possible procedures to guide the system toward pure states
characterized by fixed correlation conditions, on the other hand
suggests a way to to estimate the coupling strenght between each
of the $N$ spins and the central one.

\section{Disordered spin star system}
\begin{figure}\begin{center}\resizebox{0.4\columnwidth}{!}{\includegraphics{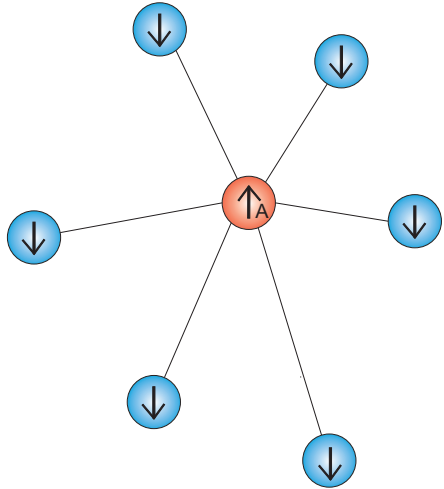}}
\caption{An illustration of a disordered spin star system
}\end{center}\end{figure}

Having in mind as objective the possibility of generating fixed
entanglement conditions in a system of $N$ uncoupled spins
$\frac{1}{2}$, we exploit the interaction of each element of the
system with another spin $\frac{1}{2}$ hereafter called central
one. The system we are talking about is illustrated for
convenience in Figure 1 and it can be described adopting the
following hamiltonian model:
\begin{equation}\label{Hamiltonian1}
H=H_0+H_I
\end{equation}
with
\begin{equation}\label{Hamiltonian2}
H_0=\omega\sum_{j=1}^N\sigma_z^j+\omega_0\sigma_z^A,
\end{equation}
\begin{equation}\label{Hamiltonian3}
H_I=\sum_{j=1}^N\alpha_j(\sigma_+^A\sigma_-^j+\sigma_-^A\sigma_+^j),
\end{equation}
The Pauli operators labelled by the index $A$ refer to the central
spin, while the others characterized by the index $j$ ($j=1..N$)
refer to the $N$ spins. The coupling strength between the spin $j$
and the central one is measured by the constant $\alpha_j$ and,
generally speaking, $\alpha_j$ may be  different from $\alpha_i$
for $j\neq i$. In realistic physical situations the coupling
constants can change for example proportionally to the distance
from the central system \cite{Braktavatsala}. We refer to the
system described by eq. (\ref{Hamiltonian1}) as \textit{disordered
spin star system}.

Studying the dynamical properties of this model is for example of
interest in contexts like quantum dot coupled by hyperfine
interaction with nuclear spins, or electronic spins bound to
phosphorus atoms in a matrix of silica or germanio in presence of
defects \cite{Schliemann}.

The symmetry properties of the hamiltonian model can be
successfully exploited in order to analyze the dynamics of the
system. It is easy to convince oneself that the component along
the $z$ axes of the total angular momentum operator
$S_z=\frac{\sigma_z^A}{2}+\frac{1}{2}\sum_{j=1}^N\sigma_z^j=\frac{\sigma_z^A}{2}+J_z$
is a constant of motion. Then, starting from an eigenstate of
$S_z$ the system evolves in the correspondent invariant Hilbert
subspace. Let's suppose in particular to prepare the system in the
following state
\begin{equation}\label{condizione}
|\psi_{\{i_1,...i_p\}}(0)\rangle=|\!\!\uparrow_A\rangle|\!\!\downarrow\ldots\uparrow_{i_1}
\ldots\uparrow_{i_p}\ldots\downarrow\rangle,
\end{equation}
where the central spin, as well as $p$ of the $N$ spins, namely
the $i_1$-th, $i_2$-th,... $i_p$th, are in their respective up
state $\vert\!\!\uparrow\rangle$ defined as $\sigma_z^i
\vert\!\!\uparrow\rangle=\vert\!\!\uparrow\rangle$ with $i=A,
1,..,N$, whereas the others are in their down state
$\vert\!\!\downarrow\rangle$ with $\sigma_z^i
\vert\!\!\downarrow\rangle=-\vert\!\!\downarrow\rangle$.

We in addition denote the state
$|\!\!\uparrow_A\rangle|\!\!\downarrow,\downarrow,\ldots,\downarrow\rangle$,
where all the uncoupled spins are down, by $|\psi_0(0)\rangle$.

 Taking into account the previous considerations
we may claim that, at time instant $t$ the state of the system
prepared in the state (\ref{condizione}) can be written as
\begin{eqnarray}\label{psi(t)}\nonumber
|\psi_{\{i_1,...i_p\}}(t)\rangle=&\sum_{j_1<j_2<\ldots<j_p}a_{j_1,j_2,\ldots,j_p}(t)|\!\!\uparrow_A\rangle|\!\!\downarrow\ldots\uparrow_{j_1}
\ldots\uparrow_{j_p}\ldots\downarrow\rangle+\\
&+\sum_{j_1<j_2<\ldots<j_{p+1}}b_{j_1,j_2,\ldots,j_{p+1}}(t)|\!\!\downarrow_A\rangle|\!\!\downarrow\ldots\uparrow_{j_1}
\ldots\downarrow\ldots\uparrow_{j_{p+1}}...\downarrow\rangle,
\end{eqnarray}

In equation (\ref{psi(t)}) each index $j_i$ $(i=1,..,p+1)$ runs
from 1 to $N$. Thus the first term of the right hand side is a
superposition of all the states in which the central spin, as well
as $p$ among the $N$ uncoupled spins, are in their up state. The
second term of equation (\ref{psi(t)}) is similarly a linear
combination of all the states in which the central system is in
its down state whereas $p+1$ of the $N$ spins are in their up
state. We notice that the structure of $\vert \psi_0(t)\rangle$
can be deduced from that of $|\psi_{\{i_1,...i_p\}}(t)\rangle$
simply substituting to the first sum of eq. (\ref{psi(t)}) the
term $a(t)|\!\!\uparrow_A\rangle\\|\!\!\downarrow\ldots\downarrow
\ldots\downarrow\rangle$. Inserting eq. (\ref{psi(t)}) in the
time-dependent Schr\"{o}dinger equation leads to the following
system of coupled equations for the probability amplitudes
$a_{j_1,j_2,\ldots,j_p}(t)$ and $b_{j_1,j_2,\ldots,j_{p+1}}(t)$
\begin{eqnarray}\label{a1}
i\,\dot{a}_{j_1,j_2,\ldots,j_p}(t)&=&\Delta\,a_{j_1,j_2,\ldots,j_p}(t)+
\sum_{r=1(r\neq j_1,..j_p)}^N\alpha_rb_{O(\{j_1,\ldots,j_p\}\cup\{r\})}(t)\\
\label{b1}i\,\dot{b}_{j_1,j_2,\ldots,j_{p+1}}(t)&=&-\Delta\,
b_{j_1,j_2,\ldots,j_{p+1}}(t)+\sum_{r=1(r\in
\{j_1,..j_p\})}^N\alpha_ra_{\delta_r(j_1,\ldots,j_{p+1})}(t)\end{eqnarray}
where $\Delta=\omega-\omega_0$. In eq.(\ref{a1}) we have
introduced the operator $O$ which adds the index $r$ to the set of
indices $\{j_1,\ldots,j_p\}$ arranging them in increasing order.
We point out that this operator is well defined if $r$ does not
belong to the set $\{j_1,\ldots,j_p\}$, otherwise the probability
amplitude $b_{O(\{j_1,\ldots,j_p\}\cup\{r\})}(t)$ would have $p$
indices instead of $(p+1)$ becoming senseless. The operator
$\delta_r$ appearing in turn in eq.(\ref{b1}) acts on the family
of $p+1$ indices, recovering a set of $p$ indices from
$\{j_1,j_2,\ldots,j_{p+1}\}$ by eliminating the index $r$. We have
to mention that the above operator is well defined if $r$ belongs
to the set $\{j_1,j_2,\ldots,j_{p+1}\}$ in order to assure the
correct definition of a probability amplitudes of the type
$a_{j_1,j_2,\ldots,j_p}(t)$.

This system of differential equations can be easily decoupled when
the system is prepared in the state $\vert \psi_0(0)\rangle$. In
this case eqs. (\ref{a1}) and (\ref{b1}) become
\begin{equation}\label{a1_0}
i\,\dot{a}(t)=\Delta\,a(t)+\sum_{j=1}^N\alpha_jb_{j}(t)
\end{equation}
\begin{equation}\label{b1_0}
i\,\dot{b}_j(t)=-\Delta\, b_j(t)+\alpha_ja(t),
\end{equation}
and it is easy to show  that $a(t)$ fulfills the following Cauchy
problem
\begin{eqnarray}\label{Cauchy}
&&a(0)=1\\
&& \ddot{a}(t)=-\left(\Delta^2+\sum_{r=1}^N\alpha_r^2\right)a(t),
\end{eqnarray}
whose solution is
\begin{equation}\label{a4}
a(t)=\cos\left(\sqrt{\sum_{j=1}^N\alpha_j^2+\Delta^2}\;t\right)-i\frac{\Delta}{\sqrt{\sum_{j=1}^N\alpha_j^2+\Delta^2}}\sin\left(\sqrt{\sum_{j=1}^N\alpha_j^2+\Delta^2}\;t\right)
\end{equation}
Inserting eq. (\ref{a4}) in eq. (\ref{b1_0}) leads to a first
order non homogeneous linear differential equation for $b_j(t)$
which can be easily solved from the initial conditions $b_j(0)=0$
getting
\begin{equation}\label{b4}
b_j(t)=-i\frac{\alpha_j}{\sqrt{\sum_{j=1}^N\alpha_j^2+\Delta^2}}\sin\left(\sqrt{\sum_{j=1}^N\alpha_j^2+\Delta^2}\;t\right).
\end{equation}
When the system is prepared in the state (\ref{condizione}), eqs.
(\ref{a1}) and (\ref{b1}) may be still managed in such a way to
decouple the probability amplitudes of the type
$a_{j_1,j_2,\ldots,j_p}$ from those of type
$b_{j_1,j_2,\ldots,j_{p+1}}$. We do not present here such a
procedure since in what follows we concentrate on the rich
dynamical properties of the system evolving in accordance with
$\vert \psi_0(t)\rangle$.

\section{Generation of W-like states}
The results obtained in the previous Section suggest that
measuring the central spin we have the possibility of guiding the
system of interest, namely the $N$ uncoupled spins, toward a
linear coherent superposition of states characterized by the fact
that only one spin is in the state $|\!\!\uparrow\rangle$ whereas
the others are in the state $|\!\!\downarrow\rangle$.

Starting from eq.(\ref{psi(t)})  we may indeed claim that a
measure of the observable $\sigma_z^A$ gives, with probabilities
$\sum_{j=1}^N|b_j(t)|^2$, the eigenvalue $-1$. In this case the
$N$ spins are left in the normalized state
\begin{equation}\label{Stato1}
|W^{gen}\rangle=\frac{1}{\sqrt{\sum_{j=1}^N\alpha_j^2+\Delta^2}}\left(\alpha_1|\!\!\uparrow,\downarrow,\ldots,\downarrow\rangle+\ldots+\alpha_j|\!\!\downarrow,\ldots,\uparrow,\ldots,\downarrow\rangle+\ldots+\alpha_N|\!\!\downarrow,\downarrow,\ldots,\uparrow\rangle\right).
\end{equation}
The state given by eq.(\ref{Stato1}) looks like the well known
W-state \cite{Horodecki}
\begin{equation}\label{Stato2}
|W\rangle=\frac{1}{\sqrt{N}}\left(|\!\!\uparrow,\downarrow,\ldots,\downarrow\rangle+\ldots+
|\!\!\downarrow,\ldots,\uparrow,\ldots,\downarrow\rangle+\ldots+|\!\!\downarrow,\downarrow,\ldots,\uparrow\rangle\right).
\end{equation}
the only difference between the two states (\ref{Stato1}) and
(\ref{Stato2}) being the weight of each component in the
superposition. In the W-state  all the states of the superposition
appear indeed with the same probability. On the other hand the
state (\ref{Stato1}) we have obtained with our procedure, reduces
to the W-state when the central spin does not distinguish the $N$
spins around it, that is when $\alpha_j\equiv\alpha\;\forall
j=1,\ldots,N$. For these reasons we
 call the state $|W^{gen}\rangle$ a W-like state.

It is important to underline that the procedure we have discussed
is a conditional one. In other words we may claim to generate the
state $|W^{gen}\rangle$ only if the measurement of $\sigma_z^A$
gives the eigenvalue $-1$. Starting from eq. (\ref{b4}) we can
write the probability of success of our procedure as follows
\begin{equation}\label{probabilita}
P=\sum_{j=1}^N\frac{\alpha_j^2}{\sum_{j=1}^N\alpha_j^2+\Delta^2}\sin^2\left(\sqrt{\sum_{j=1}^N\alpha_j^2+\Delta^2}\;t\right).
\end{equation}
Thus, appropriately choosing the time instant $t$ at which the
measure of $\sigma_z^A$ is performed, we may generate the desired
state with the highest probability that coincides with one in
correspondence to $\Delta=0$. If indeed we measure the observable
$\sigma_z^A$ at time instant
\begin{equation}\label{tn}
t_n=\frac{\pi(2n+1)}{2\sqrt{\sum_{j=1}^N\alpha_j^2+\Delta^2}},\quad
n=0,1,2,\ldots
\end{equation}
the probability of success becomes
$P=\sum_{j=1}^N\frac{\alpha_j^2}{\sum_{j=1}^N\alpha_j^2+\Delta^2}$
and thus $P=1$ if $\Delta=0$.

It is on the other hand of relevance to analyze the behaviour of
such a probability with respect to imprecisions in setting the
time instant at which the measurement of the observable
$\sigma_z^A$ is performed. Let's first of all observe that the
probability of success $P$, as given by equation
(\ref{probabilita}), is a periodic function of $t$, the period
being $T=\pi(\sum_{j=1}^N\alpha_j^2+\Delta^2)^{-\frac{1}{2}}$.
This circumstance suggests to choice the time instant at which to
perform the measurement optimizing, as far as possible, our
proposal on the experimental side too. To this end let's consider
for simplicity the case $\Delta=0$ and indicate by
$t_n^*=t_n+\delta$ the  time instant at which the measurement if
performed. If $\delta$ is small enough the probability of success
does not appreciably get reduced. Let's suppose in particular that
$|\delta| \leq T 10^{-1}$.  Moreover it is also reasonable to
assume that the relative error $\frac{\delta}{t_n}$ is of the
order of $10^{-3}$ thus implying that
\begin{equation}\label{magnitude tn}
t_n\sim10^3\delta\leq 10^2\frac{\pi}{\sqrt{\sum_{j=1}^N\alpha_j^2}}
\end{equation}
If our model describes, for example, hyperfine interaction of a
localized electron with nuclei, the order of magnitude of the
coupling constant $\alpha_j$ can be estimated as $10^{-5} eV$
\cite{Merkulov}, \cite{Mattis} in correspondence to which $t_n$
must be less than or equal to $\pi 10^{-9}sec$. Thus $t_n\sim 1
ns$ is compatible with eq. (\ref{magnitude tn}) and it may be
realized fixing $n=100$.
\begin{figure}[h]\begin{center}
\resizebox{0.70\columnwidth}{!}{\includegraphics{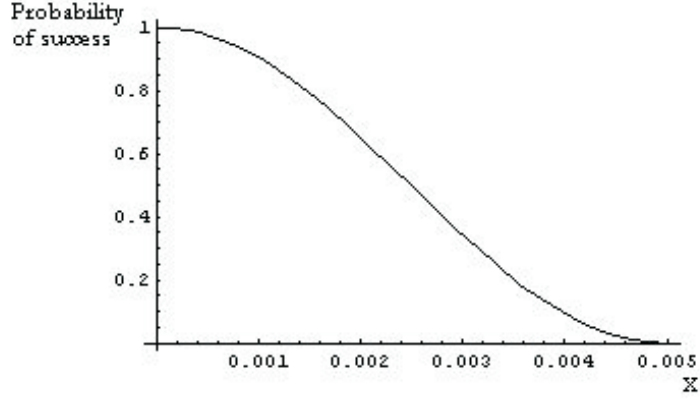} }
\caption{Probability of success to generate $|W^{gen}\rangle$ as
function of $x=\frac{\delta}{t_{n}}$ in correspondence to
$\Delta=0$ and n=100.} \label{fig2}
\end{center}\end{figure}
In figure 2 we plot the probability of success of our scheme given
by eq. (\ref{probabilita}) versus $x=\frac{\delta}{t_{100}}$
putting $\Delta=0$ and $t=t_{100}+\delta$. As foreseeable, this
probability of success remains of experimental interest for $x$ up
to $3\cdot 10^{-3}$. Thus we may conclude that the procedure is
stable enough against unavoidable uncertainties in the time
instant at which the measurement of $\sigma_z^A$ is done.

Before concluding this section it is interesting to emphasize that
the coupling between the $N$ spins and the central system
generates entanglement between any two spins around the central
one. In order to estimate the amount of such an entanglement and
analyze its time evolution we evaluate the relative concurrence
function. It is easy to prove that, starting from the initial
state $\vert \psi_0(0)\rangle$, the reduced density $\rho_{ij}(t)$
describing the system of the two spins $i$ and $j$ among the $N$
uncoupled ones, can be written in the form
\begin{equation} \label{roij}
\rho_{ij}(t)=\left(%
\begin{array}{cccc}
  0 & 0 & 0 & 0 \\
  0 & |b_i(t)|^2 & b_i(t)b_j(t)^{\ast} & 0 \\
  0 & b_i(t)^{\ast}b_j(t) & |b_j(t)|^2 & 0 \\
  0 & 0 & 0 & |a(t)|^2+\sum_{k\neq i,j}|b_k(t)|^2 \\
\end{array}%
\right)
\end{equation}
when expressed in the standard two-spin basis
$\{|\!\!\uparrow\uparrow\rangle,|\!\!\uparrow\downarrow\rangle,
|\!\!\downarrow\uparrow\rangle,|\!\!\downarrow\downarrow\rangle\}$.
In equation (\ref{roij})$b_j(t)$ and $a(t)$ are given by eqs.
(\ref{b4}) and (\ref{a4}) respectively. Thus the concurrence
function $C_{ij}(t)$ turns out to be
\begin{equation}\label{cij}
C_{ij}(t)=\frac{2|\alpha_i||\alpha_j|}
{\sum_{j=1}^N\alpha_j^2+\Delta^2}\sin^2\left(\sqrt{\sum_{j=1}^N\alpha_j^2+\Delta^2}\;t\right)
\end{equation}
As expected, the degree of entanglement get established between
the two uncoupled spins $i$ and $j$, oscillates with time $t$. In
addition the maximum value of $C_{ij}(t)$ is proportional to
$|\alpha_i||\alpha_j|$ and when $\alpha_k=\alpha$ $\forall k$ and
$\Delta=0$ reaches the value $\frac{2}{N}$. This means that,
increasing the number of spins around the central one, leads to a
weak and weak pair quantum correlation in the system of interest.

\section{Generation of well-defined angular momentum states of the N uncoupled spins}
The W-state in eq.(\ref{Stato2}) is a multipartite entangled state
of the $N$ uncoupled spins around the central one. On the other
hand it coincides with a particular superposition of states in
which $N-1$ spins have projection down while only one has
projection up. This state can be thus obtained applying the
collective ladder operator $\sigma_+=\sum_i \sigma_+^{(i)}$ on the
state $|\!\!\downarrow,\downarrow,\ldots,\downarrow\rangle$
 that is $|W\rangle\propto\sum_{i=1}^N\sigma_+^{(i)}|\!\!\downarrow,\downarrow,\ldots,\downarrow\rangle$.
In other words the W-state is a common eigenstate of
$\mathbf{J}^2$ and $J_z$, $\mathbf{J}=\frac{1}{2}\sum_{i=1}^N
\mathbf{\sigma}_i$ being the collective angular momentum operator
of the $N$ uncoupled spins, with
\begin{eqnarray}
&\mathbf{J}^2\vert W\rangle=\frac{N}{2}(\frac{N}{2}+1)\vert
W\rangle\\
&J_z\vert W\rangle=(-\frac{N}{2}+1)\vert W\rangle
\end{eqnarray}
Thus $\vert W \rangle\equiv|J,M\rangle$ whit $J=\frac{N}{2}$ and
$M=-\frac{N}{2}+1$.

This observation suggests us the possibility to iterate our
procedure in order to generate all the well defined angular momentum
states $|J=\frac{N}{2},M\rangle$ with
$M=-\frac{N}{2}\ldots\frac{N}{2}$.

Let's indeed consider the spin star system in which the central
spin interacts in the same way with all the others uncoupled $N$
spins, that is $\alpha_j\equiv\alpha\;\forall j$. Under this
condition the hamiltonian model (\ref{Hamiltonian1}) is invariant
by permutation of an arbitrary couple of spins among the $N$.
Moreover
$[\sigma^{A\,2},H]=[J^2,H]=[\sigma_z^A+J_z,H]=[J^2_{int},H]=0$,
$\vec{J}_{int}$ being an intermediate angular momentum resulting
from the coupling of selected at will individual angular momentum
of the $N$ spins. These symmetry properties suggest to develop the
dynamics of our system exploiting the coupled angular momentum
basis $\{|J,M,\nu\rangle\}$ for the $N$ spins instead of the
factorized one previously used. The index $\nu$ runs from 1 to
$\nu_{MAX}(J)$ and allows us to distinguish between different
states of the basis characterized by the same $J$ and $M$.

In the initial condition  $|\psi_0(0)\rangle$ the $N$ spins around
the central one are in the coupled angular momentum state
$|J=\frac{N}{2},M=-\frac{N}{2},\nu=1\rangle$. In what follows we
do not indicate anymore the the index $\nu$ remaining it equal to
one. Thanks to the symmetry properties of our system at a generic
time instant $t$ we can rewrite the state of the total system in
the form
\begin{equation}\label{generazione2}
|\psi(t)\rangle=A_1(t)|\uparrow_A\rangle\vert\frac{N}{2},-\frac{N}{2}\rangle+B_1(t)|\downarrow_A\rangle\vert\frac{N}{2},-\frac{N}{2}+1\rangle
\end{equation}
where
\begin{equation}
A_1=\cos(p_1\alpha t),
\end{equation}
\begin{equation}
B_1=-i\sin(p_1\alpha t),
\end{equation}
with
\begin{equation}\label{p}
p_1=\sqrt{\frac{N}{2}\left(\frac{N}{2}+1\right)-
\left(-\frac{N}{2}\right)\left(-\frac{N}{2}+1\right)}\equiv\sqrt{N}.
\end{equation}

Thus as before, if we assume the central spin $A$ measured in the
state $\vert \downarrow_A\rangle$ the $N$ uncoupled spins are
projected onto the W-state $\vert W \rangle\equiv \vert
\frac{N}{2},-\frac{N}{2}+1\rangle$. As demonstrated in the
previous Section the probability of success to generate the
W-state coincides with $\sin^2(p_1\alpha t)$ and it is equal to 1
if the measurement is performed at time instants
$t_n=\frac{\pi(2n+1)}{2\alpha p_1}$ obtained from equation
(\ref{tn}) putting $\alpha_j\equiv\alpha$ $\forall j$ and
$\Delta=0$. Suppose now to iterate the procedure preparing once
again the central spin in the up state. The new initial condition
is then
\begin{equation}\label{psi02}
\vert \psi(0)\rangle=\vert \uparrow_A\rangle \vert
\frac{N}{2},-\frac{N}{2}+1\rangle
\end{equation}
that as easily demonstrable evolves as
\begin{equation}\label{psit2}
\vert \psi(t)\rangle=A_2(t)\vert \uparrow_A\rangle \vert
\frac{N}{2},-\frac{N}{2}+1\rangle+B_2(t)\vert \downarrow_A\rangle
\vert \frac{N}{2},-\frac{N}{2}+2\rangle
\end{equation}
with
\begin{eqnarray}
A_2(t)=\cos(p_2\alpha t)\\
B_2(t)=-i\sin(p_2\alpha t)
\end{eqnarray}
where
$p_2=\sqrt{\frac{N}{2}(\frac{N}{2}+1)-(-\frac{N}{2}+1)(-\frac{N}{2}+2)}$.
Eq. (\ref{psit2}) immediately implies that, measuring the spin $A$
in its down state, makes the $N$ spins to collapse onto the
angular momentum state $\vert \frac{N}{2},-\frac{N}{2}+2\rangle$.
It is possible at this point to convince oneself that, iterating
$k$-times our procedure, we generate the coupled angular momentum
state $\vert \frac{N}{2},-\frac{N}{2}+k\rangle$. As far as the
probability of success $P_k$ that after $k $ measurements the $N$
uncoupled spins are left in the state $\vert
\frac{N}{2},-\frac{N}{2}+k\rangle$, it is easy to prove that it is
given by
\begin{equation}\label{Pk}
P_k=\prod_{i=1}^k \sin^2(p_i \alpha t)
\end{equation}
where
$p_i=\sqrt{\frac{N}{2}(\frac{N}{2}+1)-(-\frac{N}{2}+i-1)(-\frac{N}{2}+i)}$.

Thus, if at the $i-$th step we have the possibility of choosing the
time instant at which performing the measurement act on the central
spin $A$ in such a way that $t_n^{(i)}=\frac{(2n+1)\pi}{2\alpha
p_i}$, at least in principle we may claim that the desired state
$\vert \frac{N}{2},-\frac{N}{2}+k\rangle$ of the $N$ spins are
generated with certainty.

\section{Estimating the order of magnitude of the coupling constant $\alpha_j$}
As we are going to prove the dynamics of our system can be also
successfully exploited in order to estimate the coupling intensity
between each of the $N$ distinguishable and non interacting $N$
spins and the central one.

Let's indeed consider for simplicity the case $\Delta=0$ and,
under this condition, concentrate on the behaviour of the
probability $P_0(t)$ to find the system,  at a generic time
instant $t$, in the initial condition  $|\psi_0(0)\rangle$. The
results obtained in the previous Section immediately implies that
such a probability is given by
\begin{equation}\label{po}
P_0(t)=|a(t)|^2=\cos^2\left(\sqrt{\sum_{j=1}^N\alpha_j^2}\;t\right).
\end{equation}
Thus, the probability of recovering the $N$ uncoupled spins as well
as the central one in their initial state is a periodic function of
$t$, the period being in inverse relation to the quantity
$\sum_{j=1}^N\alpha_j^2$. If, as reasonable, we assume that all the
coupling constants are of the same order, analyzing the temporal
behaviour of $P_0(t)$, we have the possibility of estimating the
order of magnitude of each $\alpha_j$. It is important to stress
that the possibility of knowing at least the order of magnitude of
the coupling constants $\alpha_j$ play a central role for example in
all the cases in which the spin of an electron localized in a
quantum dot is used as realization of a quantum bit \cite{Merkulov}.
In these cases indeed the spin relaxation mechanism is mainly
connected with its interaction with bulk nuclear spins.

Let's moreover observe that the knowledge of the frequency of the
function $P_0(t)$ given by eq. (\ref{po}) can be also exploited in
order to estimate how much disordered the spin star system model
(\ref{Hamiltonian1}) is. Let's suppose indeed that the $N$ spins
of interest have been prepared in a W-like state following the
procedure previously discussed. At this point, if we measure the
observable $\sigma_z^j$ finding the eigenvalue $+1$, then the
total system is projected onto the state
$|\!\!\downarrow,\downarrow,\ldots,\uparrow_j,\ldots,\downarrow\rangle$
in which only the $j$-th spin is in its upper state whereas the
others are in their respective down state. The probability of such
an event, directly obtainable starting from eq.(\ref{Stato1}),
exactly coincides with the quantity
$\frac{\alpha_j^2}{\sum_{j=1}^N\alpha_j^2}$.

We may thus conclude that knowing the probability of success to
find the $j$-th spin in the up state $|\!\!\uparrow\rangle$ when
the system of $N$ spins is prepared in a W-like state, allows us
to give an estimation of the interaction strength between the
$j$-th spin and the central one.

\section{Conclusion}

Generally speaking the possibility of establishing on demand fixed
entanglement conditions in a multipartite system is an interesting
objective both in its own and also in view of its applicative
potentialities. In this paper in particular we have concentrated
on a multipartite system composed by $N$ not interacting spins
$\frac{1}{2}$. In order to guide this system toward assigned
entangled states, we have exploited the interaction between each
of the $N$ subsystems with a single spin $\frac{1}{2}$. The
disordered spin star system thus obtained has been successfully
used to generate W-like states as well as well-defined angular
momentum states of the $N$ uncoupled spins.

The study of the exact dynamics of the disordered spin star system
reported in this paper, has provided the possibility to envisage a
way to estimate at least the order of magnitude of the coupling
strength between each of the $N$ uncoupled spins and the central
one. To gain this information is, for example, of particular
relevance when electron spin relaxation plays an important role.
It is indeed appropriate to remark that our system can be adopted
to describe hyperfine interaction of a single electron spin with
nuclei in quantum dots and that this interaction mechanism may be
the dominant source of electron spin relaxation.

\end{document}